\documentclass[review]{elsarticle}
\usepackage{lineno,hyperref}
\usepackage{amssymb}
\usepackage{amsfonts}
\usepackage{amsmath}
\usepackage{graphicx}

\modulolinenumbers[5]
\journal{Arxiv}
\begin{document}

\begin{frontmatter}

\title{Generalized Version of the Creation and Annihilation Operators for the
Schr\"{o}dinger Equation}

\author[mysecondaryaddress]{Santos, L. C. N \corref{mycorrespondingauthor}}
\ead{luis.santos@ufsc.br}

\author[mysecondaryaddress]{C.C. Barros Jr\corref{mycorrespondingauthor}}

\cortext[mycorrespondingauthor]{Corresponding author}
\ead{barros.celso@ufsc.br}

\address[mysecondaryaddress]{Departamento de F\'{\i}sica - CFM - Universidade Federal de Santa
Catarina, Florian\'{o}polis - SC - CP. 476 - CEP 88.040 - 900 - Brazil.}

\begin{abstract}
A generalized version of the creation and annihilation operators is constructed and the factorization of the Schr\"{o}dinger equation is investigated. It is shown
that the generalized version of factorization operators yield a factorization
for the twelve different separable coordinates for the Schr\"{o}dinger equation.
\end{abstract}

\begin{keyword}
  \sep factorization method \sep Schr\"{o}dinger equation \sep operators

\end{keyword}

\end{frontmatter}

\linenumbers

\section{Introduction}

The factorization method, introduced by Schr\"{o}dinger \cite{facto1} and
Dirac \cite{facto2} and later developed by Infeld and Hull \cite{facto3}, is
one of the methods for solving quantum mechanical problems. The idea is to
consider a pair of first-order differential equations which are equivalent
to a given second-order differential equation. The complete set of
normalized eigenfunctions can be obtained by the successive application of
the ladder operators on the eigenfunctions, which are the exact solutions of
the first order differential equation.

From the 1970s to the early 1980, it was a common opinion that the method
was completely explored. However, Mielnik made an additional contribution to
the traditional factorization method in 1984 \cite{facto4}. In that work, he
did not consider the particular, but the general solution to the Riccati
type equation connected with the Infeld-Hull approach. Mielnik factorization
is a powerful tool in the derivation of new Hamiltonians whose corresponding
eigenproblem is analytically solvable. On the other hand, the connection
between the Infeld method and supersymmetric quantum mechanics (SUSY QM)
has been explored by many authors \cite{facto5,facto6,facto7,facto8}. For
example, Witten noticed the possibility of arranging the second-order
differential equations into isospectral pairs, the so-called supersymmetric
partners.

In many works, the factorization method has been used as a tool to formulate
algebraic approaches to many non-relativistic quantum problems, the idea is
build sets of one variable radial operators which are realizations for $%
su\left( 1,1\right) $ Lie algebra \cite{facto9,facto10}. The separable
coordinate systems for the Schr\"{o}dinger equation are confocal quadric
surfaces \cite{facto11}, and the potential is a function of the coordinates 
\cite{facto12}.

In this paper, the possibility of factorization of the separated equations
is investigated, and it is shown that a generalized version of the creation
and annihilation operators can be constructed. Indeed, these operators yield
a factorization for the twelve different separable coordinates for the Schr%
\"{o}dinger equation. In the Infeld method, the original form of the
second-order differential equation \ 
\begin{equation}
\frac{d}{d\theta }\left( p\frac{d\psi }{d\theta }\right) +q\left( \theta
\right) \psi +\lambda \rho \left( \theta \right) \psi =0,  \label{eq00}
\end{equation}%
is transformed in the form 
\begin{equation}
\frac{d^{2}y}{dx^{2}}+r\left( x,m\right) y+\lambda y=0,  \label{eq00b}
\end{equation}%
where $m=1,2,3...$ and $p,\rho $ are positive functions. The transformation
connecting these equations is 
\begin{equation}
y=\left( p\rho \right) ^{1/2}\psi \text{, \ }dx=\left( \rho /p\right)
^{1/2}d\theta .  \label{eq00c}
\end{equation}%
In this work, we propose to apply a factorization method for the original
form of the separated Schr\"{o}dinger equation. In this way, our approach
yield factorization operators for the twelve different separable coordinate
systems. An interesting feature of this work is that the original Hilbert
space of theory is sustained.

So, this paper will show the following contents: In section II, we enumerate
the coordinates systems which will allow separation of the Schr\"{o}dinger
equation. In section III, \ a generalized version of the creation and
annihilation operators is proposed. In section IV and V, we apply the method
to the radial second-order Schr\"{o}dinger equation.

\section{Separable coordinate systems for the wave equation}

In this section, we study the separable coordinate systems for the Schr\"{o}%
dinger equation, for more details and information see \cite{facto13}. We
consider the standard differential equation for the scalar field 
\begin{equation}
\nabla ^{2}\psi +k_{1}^{2}\psi =0\text{, }  \label{eq1}
\end{equation}%
where $\triangledown ^{2}$ is the Laplace operator. When $k_{1}$ is a
function of the coordinates, we obtain the Schr\"{o}dinger equation. The
rectangular \ coordinates $x$, $y$, $z$ and the curvilinear coordinates $\xi
_{1}$, $\xi _{2}$, $\xi _{3}$ are related by the scale factors 
\begin{equation}
h_{n}=\sqrt{\left( \frac{\partial x}{\partial \xi _{n}}\right) ^{2}+\left( 
\frac{\partial y}{\partial \xi _{n}}\right) ^{2}+\left( \frac{\partial z}{%
\partial \xi _{n}}\right) ^{2}},  \label{eq2}
\end{equation}%
where $n=1,2,3$. The Laplacian can be expressed in its generalized form 
\begin{equation}
\nabla ^{2}\psi =\underset{n}{\sum }\frac{1}{h_{1}h_{2}h_{3}}\frac{\partial 
}{\partial \xi _{n}}\left[ \frac{h_{1}h_{2}h_{3}}{h_{n}^{2}}\frac{\partial
\psi }{\partial \xi _{n}}\right] ,  \label{eq3}
\end{equation}%
thus we can rewrite the equation $\left( \ref{eq1}\right) $ into%
\begin{equation}
\underset{n}{\sum }\frac{1}{h_{1}h_{2}h_{3}}\frac{\partial }{\partial \xi
_{n}}\left[ \frac{h_{1}h_{2}h_{3}}{h_{n}^{2}}\frac{\partial \psi }{\partial
\xi _{n}}\right] +k_{1}^{2}\psi =0\text{.}  \label{eq4}
\end{equation}

In order to obtain the separated equations, we introduce the St\"{a}ckel
determinant%
\begin{align}
S & =\left\vert \Phi_{mn}\right\vert =\left\vert 
\begin{array}{ccc}
\Phi_{11} & \Phi_{12} & \Phi_{13} \\ 
\Phi_{21} & \Phi_{22} & \Phi_{23} \\ 
\Phi_{31} & \Phi_{32} & \Phi_{33}%
\end{array}
\right\vert  \notag \\
& =\Phi_{11}\Phi_{22}\Phi_{33}+\Phi_{12}\Phi_{23}\Phi_{31}+\Phi_{13}\Phi
_{21}\Phi_{32}  \notag \\
& -\Phi_{13}\Phi_{22}\Phi_{31}-\Phi_{11}\Phi_{23}\Phi_{32}-\Phi_{12}\Phi
_{21}\Phi_{33},  \label{eq5}
\end{align}
where $\Phi_{mn}$ are functions of $\xi_{n}$ alone.

If the separated equations for the three-dimensional case are 
\begin{equation}
\frac{1}{f_{m}\left( \xi _{m}\right) }\frac{d}{d\xi _{m}}\left[ f_{m}\left(
\xi _{m}\right) \frac{dX_{m}}{d\xi _{m}}\right] +\underset{n}{\sum }\Phi
_{mn}\left( \xi _{m}\right) k_{n}^{2}X_{m}=0\text{,}  \label{eq6}
\end{equation}%
\ \bigskip we can relate the equations \ $\left( \ref{eq4}\right) $ and $%
\left( \ref{eq6}\right) $ by the Robertson condition \cite{facto13}%
\begin{equation*}
\frac{h_{1}h_{2}h_{3}}{S}=f_{1}\left( \xi _{1}\right) f_{2}\left( \xi
_{1}\right) f_{3}\left( \xi _{1}\right) ,
\end{equation*}%
which limits the kinds of coordinates systems which will allow separation.
The equation $\left( \ref{eq6}\right) $ is supposed to be a separated
equation, therefore the functions $f_{n}$, $\Phi _{n1}$, $\Phi _{n2}$, and $%
\Phi _{n3}$ must all be functions of $\xi _{n}$ alone.

\section{Creation and annihilation operators}

In this section, we propose a generalized version of the creation and
annihilation operators. For the Schr\"{o}dinger equation the constant $k_{1}$
must have the form $k_{1}^{2}=\varepsilon -\underset{n}{\overset{3}{\sum }}%
v_{n}\left( \xi _{n}\right) $ \cite{facto13}, where $v_{n}\left( \xi
_{n}\right) =\frac{2m}{\hbar ^{2}}V_{n}\left( \xi _{n}\right) $,
substituting it into the equation $\left( \ref{eq6}\right) $ gives%
\begin{equation*}
\frac{1}{f_{m}\left( \xi _{m}\right) }\frac{d}{d\xi _{m}}\left[ f_{m}\left(
\xi _{m}\right) \frac{dX_{m}}{d\xi _{m}}\right] +
\end{equation*}%
\begin{equation*}
\Phi _{m1}\left( \xi _{m}\right) \left[ \varepsilon -\underset{n}{\overset{3}%
{\sum }}v_{n}\left( \xi _{n}\right) \right] X_{m}+
\end{equation*}%
\begin{equation}
\Phi _{m2}\left( \xi _{m}\right) k_{2}^{2}X_{m}+\Phi _{m3}\left( \xi
_{m}\right) k_{3}^{2}X_{m}=0.  \label{eq7}
\end{equation}%
Introducing $k_{1}^{\prime }=\varepsilon $, $k_{2}=k_{2}^{\prime }$, $%
k_{2}=k_{2}^{\prime }$ into $\left( \ref{eq7}\right) $ we obtain%
\begin{align}
& \frac{1}{f_{m}\left( \xi _{m}\right) }\frac{d}{d\xi _{m}}\left[
f_{m}\left( \xi _{m}\right) \frac{dX_{m}}{d\xi _{m}}\right]  \notag \\
& +\underset{m\neq n}{\overset{3}{\sum }}k_{m}^{\prime 2}\Phi _{nm}\left(
\xi _{m}\right) X_{n}-v_{n}X_{n}  \notag \\
& =k_{n}^{\prime 2}\Phi _{nn}X_{n}.  \label{eq8}
\end{align}%
These are the separated equations for the three-dimensional space. Now we
propose to apply a factorization method for the separated Schr\"{o}dinger
equation $\left( \ref{eq8}\right) $. The idea is to define two ladder
operators 
\begin{align}
A^{{}}& =\frac{d}{d\xi _{m}}+\frac{1}{2f_{m}}\frac{df_{m}}{d\xi _{m}}%
-R\left( \xi _{m}\right) ,  \label{eq8a} \\
A^{+}& =\frac{d}{d\xi _{m}}+\frac{1}{2f_{m}}\frac{df_{m}}{d\xi _{m}}+R\left(
\xi _{m}\right) ,  \label{eq8b}
\end{align}%
and to show that these operators yield a factorization of the equation $%
\left( \ref{eq8}\right) $. Indeed, if we multiply $A^{{}}$by $A^{+}$, we
find 
\begin{align}
AA^{+}& =\frac{1}{f_{m}}\frac{d}{d\xi _{m}}\left( f_{m}\frac{dX_{m}}{d\xi
_{m}}\right) -\frac{1}{4f_{m}^{2}}\left( \frac{df_{m}}{d\xi _{m}}\right) ^{2}
\notag \\
& +\frac{1}{2f_{m}}\frac{d^{2}f_{m}}{d\xi _{m}^{2}}+R^{\prime }-R^{2},
\label{eq9}
\end{align}%
and comparing with equation $\left( \ref{eq8}\right) $ we find a Riccati
type equation 
\begin{equation}
R^{\prime }-R^{2}=\epsilon +\Gamma \left( \xi _{n}\right) ,  \label{eq9b}
\end{equation}

where $\epsilon $ is a constant and%
\begin{equation}
\Gamma \left( \xi _{n}\right) =\frac{1}{4f_{m}^{2}}\left( \frac{df_{m}}{d\xi
_{m}}\right) ^{2}-\frac{1}{2f_{m}}\frac{d^{2}f_{m}}{d\xi _{m}^{2}}+\underset{%
m\neq n}{\overset{3}{\sum }}k_{m}^{\prime 2}\Phi _{nm}-v_{n}.  \label{eq9c}
\end{equation}%
The occurrence of the Ricatti equation in the factorization of second-order
di-fferential equations is a typical phenomenon. Specifically, the
factorization ope-rators convert the equation $\left( \ref{eq8}\right) $ into
product of $A$ and $A^{+}$ with a extra condition, a Riccati type equation.
The explicit solution of this type of equation, in ge-neral, is not Known 
\cite{facto14}. In the following, we will obtain two particular solutions of
the Riccati equation $\left( \ref{eq9b}\right) $ in a spherical coordinate
system for the Coulomb and isotropic oscillator potentials.

\section{Application to hydrogen atom}

We want to show how to solve the radial Schr\"{o}dinger equation with
Coulomb potential. We apply the method to the radial second-order
differential equation, in this case, the spherical coordinates are denoted
by $\xi _{1}=r$, $\xi _{2}=\theta $, $\xi _{3}=\phi $ and the $f_{n}$
functions are $f_{1}=r^{2}$, $f_{2}=1-\cos ^{2}\theta $, $f_{3}=\sqrt{1-\cos
^{2}\phi }$ $.$ Therefore the $s$ matrix is given by 
\begin{equation}
S=\left( 
\begin{array}{ccc}
1 & \frac{1}{r^{2}} & 0 \\ 
0 & \frac{1}{\cos ^{2}\theta -1} & \frac{1}{\left( \cos ^{2}\theta -1\right)
^{2}} \\ 
0 & 0 & \frac{1}{\cos ^{2}\phi -1}%
\end{array}%
\right) .  \label{eq9d}
\end{equation}%
The separation constants are then required to be of the form%
\begin{equation*}
k_{1}^{2}=\varepsilon \text{, }k_{2}^{2}=-l\left( l+1\right) \text{, }%
k_{3}^{3}=m.
\end{equation*}%
So the radial Schr\"{o}dinger equation with potential $v_{n}=-K/r$ is 
\begin{equation}
\frac{d^{2}X_{1}}{dr^{2}}+\frac{2}{r}\frac{dX_{1}}{dr}-\frac{l\left(
l+1\right) }{r^{2}}+\frac{K}{r}=-\varepsilon X_{1},  \label{eq9e}
\end{equation}%
and the Ricatti equation is 
\begin{equation}
R^{\prime }-R^{2}=\epsilon -\frac{l\left( l+1\right) }{r^{2}}+\frac{K}{r},
\label{eq9f}
\end{equation}%
the particular solution for this equation is given by 
\begin{equation*}
R=\frac{l}{r}-\frac{K}{2l},
\end{equation*}%
where $\epsilon =-\frac{K^{2}}{4l^{2}}.$ The creation and annihilation
operators in $\left( \ref{eq8a}\right) $ and $\left( \ref{eq8b}\right) $ can
be written in the form%
\begin{align*}
A_{l}^{+}& =\frac{d}{dr}+\frac{1}{r}+\frac{l}{r}-\frac{K}{2l}, \\
A_{l}& =\frac{d}{dr}+\frac{1}{r}-\frac{l}{r}+\frac{K}{2l}.
\end{align*}%
The commutator of $A_{l}$ and $A_{l}^{+}$ is $r$ dependent%
\begin{equation}
\left[ A_{l}^{+},A_{l}^{{}}\right] =A_{l}^{+}A_{l}-A_{l}A_{l}^{+}=\frac{2l}{%
r^{2}}.  \label{eq10}
\end{equation}%
Indeed, the $A_{l}$ and $A_{l}^{+}$ are creation and annihilation operators.
We can prove this directly, the first step is to consider the product
between the operators $A_{l}^{+}A_{l}$ and the radial wave function $%
X_{n,l-1}$, i.e, 
\begin{align}
A_{l}^{+}A_{l}X_{n,l-1}& =\left( H_{l-1}-\frac{K^{2}}{4l^{2}}\right)
X_{n,l-1}  \notag \\
& =\left( \varepsilon _{n,l-1}-\frac{K^{2}}{4l^{2}}\right) X_{n,l-1},
\label{eq11}
\end{align}%
in a similar way%
\begin{align}
\text{ }A_{l}A_{l}^{+}X_{n,l}& =\left( H_{l}-\frac{K^{2}}{4l^{2}}\right)
X_{n,l}  \notag \\
& =\left( \varepsilon _{n,l}-\frac{K^{2}}{4l^{2}}\right) X_{n,l}\text{.}
\label{eq12}
\end{align}%
A direct calculation shows that these operators satisfy 
\begin{equation}
\left[ H_{l},A_{l}\right] =\frac{2l}{r^{2}}A_{l}  \label{eq13}
\end{equation}%
and 
\begin{equation}
\left[ H_{l-1},A_{l}^{+}\right] =-\frac{2l}{r^{2}}A_{l}^{+},  \label{eq14}
\end{equation}%
so the action of the $A_{l}$ and $H_{l}$ on the states $X_{n.l}$ is 
\begin{align}
H_{l}A_{l}X_{n.l}& =A_{l}\left( \frac{2l}{r^{2}}+H_{l}\right) X_{n.l}  \notag
\\
& =\varepsilon _{n,l-1}A_{l}X_{n.l},  \label{eq15}
\end{align}%
this result imply 
\begin{equation*}
X_{n.l-1}\propto A_{l}X_{n.l}\text{ }
\end{equation*}%
or\qquad \qquad 
\begin{equation}
X_{n.l-1}=cA_{l}X_{n.l},  \label{eq16}
\end{equation}%
where $c$ is a constant. \bigskip Equivalently%
\begin{equation}
X_{n.l}=cA_{l}^{+}X_{n.l-1}.  \label{eq17}
\end{equation}%
These results imply that the action of the operators $A_{l\text{ }}$ and $%
A_{l}^{+}$ on the states $X_{n.l}$ and $X_{n.l-1}$ is to change the quantum
number $l$. \ In order to determinate $c$, we apply $A_{l}$ to the left-hand
side of eq. $\left( \ref{eq17}\right) $ \ 
\begin{equation}
A_{l\text{ }}X_{n.l}=cA_{l\text{ }}A_{l}^{+}X_{n.l-1}  \label{eq18}
\end{equation}%
Therefore, using equation $\left( \ref{eq16}\right) $ we find 
\begin{equation}
c=\frac{1}{\sqrt{\left( \varepsilon _{n,l}-\frac{K^{2}}{4l^{2}}\right) }}.
\label{eq19}
\end{equation}

\section{Radial harmonic oscillator}

Solutions of equation $\left( \ref{eq8}\right) $ are limited to a small set
of potentials and the radial harmonic oscillator is one of the few quantum
systems where an exact and analytical solution is known. The
time-independent Schr\"{o}dinger radial equation for the isotropic
oscillator $v_{n}=kr^{2}$ reads%
\begin{equation}
\left[ \frac{d^{2}}{dr^{2}}+\frac{2}{r}\frac{d}{dr}-\frac{l\left( l+1\right) 
}{r^{2}}-kr^{2}\right] X_{1}=-\varepsilon X_{1},  \label{eq20}
\end{equation}%
for this case, the Riccati equation is 
\begin{equation}
R^{\prime }-R^{2}=\epsilon -\frac{l\left( l+1\right) }{r^{2}}-kr^{2}
\label{eq21}
\end{equation}%
the particular solution of equation $\left( \ref{eq21}\right) $ is given by%
\begin{equation}
R=\frac{l}{r}-\sqrt{kr}\text{, \ \ \ \ \ }\epsilon =\sqrt{k}\left(
2l-1\right) .  \label{eq22}
\end{equation}%
This solution $R$ leads to factorizing operators 
\begin{equation}
A_{l}=\frac{d}{dr}+\frac{1}{r}-\frac{l}{r}+\sqrt{k}r\text{, \ \ \ }
\label{eq23}
\end{equation}%
and%
\begin{equation}
A_{l}^{+}=\frac{d}{dr}+\frac{1}{r}+\frac{l}{r}-\sqrt{k}r.  \label{eq24}
\end{equation}%
Hence, we get the following commutation rules%
\begin{equation*}
\left[ A_{l}^{+},A_{l}\right] =\frac{2l}{r}+2\sqrt{k}
\end{equation*}%
and%
\begin{align}
\left[ H_{l},A_{l}\right] & =A_{l}\left[ A_{l}^{+},A_{l}\right]   \notag \\
& =A_{l}\left( \frac{2l}{r}+2\sqrt{k}\right) \text{\ , }  \label{eq25a}
\end{align}

\begin{align}
\text{ \ \ }\left[ H_{l-1},A_{l}^{+}\right] & =A_{l}^{+}\left[
A_{l}^{{}},A_{l}^{+}\right]  \notag \\
& =-A_{l}^{+}\left( \frac{2l}{r}+2\sqrt{k}\right) ,  \label{eq25c}
\end{align}

thus, the action of the $A_{l}$ and $H_{l}$ on the states $X_{n.l}$ is%
\begin{equation}
H_{l}A_{l}X_{n,l}=\varepsilon _{n,l-1}A_{l}X_{n,l}.  \label{eq26}
\end{equation}%
Equivalently, the action of the $A_{l}^{+}$ and $H_{l-1}$ on the states $%
X_{n.l-1}$ leads to%
\begin{equation}
H_{l-1}A_{l}^{+}X_{n.l-1}=\varepsilon _{n,l}A_{l}^{+}X_{n,l-1},  \label{eq27}
\end{equation}%
thus, $A_{l}^{+}$ and $A_{l}$ are the raising and lowering operators for the
isotropic oscillator. It follows that%
\begin{equation}
X_{nl-1}=cA_{l}X_{nl}\text{, \ \ }X_{nl}=cA_{l}^{+}X_{nl-1},  \label{eq28}
\end{equation}%
where%
\begin{equation*}
c=\frac{1}{\sqrt{\varepsilon _{nl}+\sqrt{k}\left( 2l-1\right) }}.
\end{equation*}

\section{Summary and conclusions}

In this paper, we have determined the separable coordinate systems for the
Schr\"{o}dinger equation. From these results, factorization operators \ for
the twelve different separable coordinates have been provided. We have
determined the Schr\"{o}dinger equation in the presence of a Coulomb
potential and a radial harmonic oscillator potential. We have shown that, a new set of generalized creation and
annihilation operators has been introduced. By using the apparatus developed
in this work, we believe that other potentials in different coordinate
systems can be solved. 
We conclude by mentioning that links between supersymmetric quantum mechanics and non-linear ordinary differential equations as the Riccati equation $\left( \ref{eq9b}\right) $ can be established \cite{facto15}. 

\section*{References}

\bibliographystyle{model1-num-names}
\bibliography{C:/referencias/referencias_unificadas}

\end{document}